\documentclass[12pt]{elsart}

\def\Tr{{\rm Tr}}
\def\hph{{\hat\Phi}}

\begin{document}

\begin{frontmatter}
\begin{flushright}
IFUP-TH/2002-11
\end{flushright}
\title{Abelian projection in $SU(N)$ gauge theories}
\author[Pisa]{L. Del Debbio\thanksref{lddmail}},
\author[Pisa]{A. Di Giacomo\thanksref{adgmail}},
\author[Oxford]{B. Lucini\thanksref{blmail}},
\author[Pisa]{G. Paffuti\thanksref{pafmail}}

\thanks[lddmail]{ldd@df.unipi.it}
\thanks[adgmail]{adriano.digiacomo@df.unipi.it}
\thanks[blmail]{lucini@thphys.ox.ac.uk}
\thanks[pafmail]{paffuti@df.unipi.it}

\address[Pisa] {Dipartimento di Fisica, Universit\`a di Pisa,
and INFN Sezione di Pisa, Italy}
\address[Oxford] {Theoretical Physics, University of Oxford, UK}

\begin{abstract}
The abelian projection in $SU(N)$ gauge theories is discussed in
detail, as well as the construction of a disorder parameter to study
dual superconductivity as a mechanism for color confinement. If the
ideas of the large $N$ limit are correct, a universal $N$-independent
behavior is expected for the suitable rescaled disorder parameter as a
function of $\lambda=g^2 N$. 
\end{abstract}

\end{frontmatter}

\section{Introduction}
There exists evidence from lattice simulations~\cite{1,2,3} that color
confinement is produced by condensation of magnetic monopoles in the
vacuum, i.e. dual superconductivity. Vacuum behaves as a dual
superconductor in the confining phase, and goes to normal at the
deconfining phase transition.

The evidence refers to $SU(2)$ and $SU(3)$ pure gauge theories.
Preliminary data indicate that the same mechanism is at work in QCD with
dynamical quarks~\cite{4}, a fact which is in line with the ideas of
$N_c\to\infty$~\cite{5,6}. As $N_c\to\infty$ with $\lambda = g^2 N_c$
fixed the theory should preserve its structure: corrections ${\mathcal
O}(1/N_c)$ are expected to be small and under control. If this is true
the mechanism of confinement should be the same at all values of $N_c$
and also in full QCD, quark loops being non-leading in the $1/N_c$
expansion.

A direct check of these ideas can be done by exploring, by the same
techniques used in Ref.~\cite{1,3}, the symmetry of the confining
vacuum in $SU(N)$ theories. The technique used there was to measure
the vacuum expectation value $\langle\mu\rangle$ of an operator $\mu$,
which creates a monopole as a function of the temperature, across the
deconfining phase transition. A non zero $\langle\mu\rangle$ implies
dual superconductivity. This is exactly what is found below the
deconfining temperature $T_c$. Above $T_c$, $\langle\mu\rangle$
vanishes, and sectors with different magnetic charge are
superselected.

Monopoles are defined by a procedure called abelian projection~\cite{7},
which we summarize in the case of the $SU(2)$ gauge theory.

Let $\phi = \vec\phi \,\cdot\, \vec\sigma/2$ be any field in the adjoint
representation, and $\hat\phi \equiv \vec\phi/|\vec\phi|$ its direction
in color space, which is defined everywhere in a configuration except at
zeros of $\vec \phi$.

A gauge invariant, color singlet field strength tensor can be defined:
\begin{equation}
F_{\mu\nu} = \hat\phi\cdot\vec G_{\mu\nu} - \frac{1}{g}
\hat\phi\cdot(D_\mu \hat\phi\wedge D_\nu \hat\phi)\label{eq1}
\end{equation}
where
\[ \vec G_{\mu\nu} = \partial_\mu\vec A_\nu - \partial_\nu\vec A_\mu
+g \vec A_\mu\wedge \vec A_\nu\]
is the color field strength tensor, and
\[ D_\mu \hat\phi = (\partial_\mu - g \vec A_\mu\wedge)\hat\phi\]
is the covariant derivative of $\hat\phi$. The two terms in
Eq.~(\ref{eq1}) are separately color singlets and gauge invariant. The
combination is chosen in such a way that bilinear terms in $\vec A_\mu
\vec A_\nu$, and $\vec A_\mu \partial_\nu\hat\phi$ cancel.  Indeed, by
explicit computation:
\begin{equation}
F_{\mu\nu} = \partial_\mu(\hat\phi\cdot \vec A_\nu) -
\partial_\nu(\hat\phi\cdot \vec A_\mu) -\frac{1}{g}
\hat\phi\cdot(\partial_\mu\hat\phi\wedge\partial_\nu\hat\phi)
\label{eq2}
\end{equation}
If we gauge transform to make $\hat\phi = {\rm const}$, e.g.
$\hat\phi = (0,0,1)$ then the second term in Eq.~(\ref{eq2}) vanishes
and $F_{\mu\nu}$ becomes abelian
\begin{equation}
F_{\mu\nu} = \partial_\mu A^3_\nu -
\partial_\nu A^3_\mu
\label{eq4}
\end{equation}
This gauge transformation is known as abelian projection on $\hat\phi$,
and is defined up to a residual $U(1)$, corresponding to rotations around
$\hat\phi$.

Monopoles can be present in a noncompact formulation of the theory
when $\partial^\mu F^*_{\mu\nu}$ can be different from
zero~\cite{8}. A non zero magnetic current exists
\begin{equation}
j_\nu = \partial^\mu F^*_{\mu\nu}
\label{eq5}
\end{equation}
which is identically conserved
\begin{equation}
\partial^\nu j_\nu = 0
\label{eq6}
\end{equation}
The corresponding $U(1)$ symmetry can be either realized \`a la Wigner
or Higgs broken. In the first case the Hilbert space is made of
superselected sectors with definite magnetic charges. In the second
case, under very general assumptions, the vacuum behaves as a dual
superconductor.

The detection of dual superconductivity has been successfully
performed for $SU(2)$ and $SU(3)$ gauge theories ~\cite{1,2,3}. In this
paper we want to analyze the abelian projection and the construction
of the disorder parameter $\mu$ for generic $SU(N)$.

\section{Abelian projection in $SU(N)$ gauge theory}

In analogy to the construction for $SU(2)$, for an arbitrary operator
$\phi$ in the adjoint representation
\[ \phi = \sum\phi^a T^a\]
we can define a field strength tensor $F_{\mu\nu}$. Here $T^a$ $(a =
1\ldots N^2-1)$ are the generators of $SU(N)$ in the fundamental
representation, with normalization
\[ {\rm Tr}\left[ T^a T^b\right] = \frac{1}{2}\delta^{ab}\]
We write
\begin{equation}
F_{\mu\nu} = {\rm Tr}\left\{\phi G_{\mu\nu}\right\}
-\frac{i}{g}{\rm Tr}\left\{\phi[D_\mu\phi,D_\nu\phi]\right\}
\label{eq7}
\end{equation}
The normalization of $\phi$ has been left indeterminate. A change in the
normalization reflects in a change of the relative coefficients of the
two terms in Eq.~(\ref{eq7}). The notation is the usual one
\begin{eqnarray*}
A_\mu &=& A_\mu^a T^a\\
G_{\mu\nu} &=& \partial_\mu A_\nu - \partial_\nu A_\mu + i g
\left[ A_\mu,A_\nu\right]\\
D_\mu\phi &=& \partial_\mu\phi - ig \left[A_\mu,\phi\right]
\end{eqnarray*}
We want to investigate for what choice of $\phi$, if any, bilinear
terms in $A_\mu A_\nu$, $A_\mu \partial_\nu\phi$ cancel in
Eq.~(\ref{eq7}).  If this happens
\begin{equation}
F_{\mu\nu} = c\left\{{\rm Tr}\left\{\phi
(\partial_\mu A_\nu - \partial_\nu A_\mu)
\right\}
-\frac{i}{g}{\rm
Tr}\left\{\phi[\partial_\mu\phi,\partial_\nu\phi]\right\}
\right\}
\label{eq8}
\end{equation}
and $F_{\mu\nu}$ becomes abelian in the gauge where $\phi$ is diagonal,
\begin{equation}
F_{\mu\nu} = c{\rm Tr}\left\{
\partial_\mu (\phi A_\nu) - \partial_\nu (\phi A_\mu)
\right\}
\label{eq9}
\end{equation}

The condition for the cancellation of bilinear terms $A_\mu A_\nu$ in
Eq.~(\ref{eq7}), for any $A_\mu A_\nu$, reads
\begin{equation}
X_2(A_\mu,\phi) \equiv
{\rm Tr}\left(
\phi[A_\mu, A_\nu]\right) +
{\rm Tr}\left(\phi\left[[A_\mu,\phi],[A_\nu,\phi]\right]\right) = 0
\label{eq10}
\end{equation}
If $\phi_0$ is a solution of Eq.~(\ref{eq10}), then 
\begin{equation}
\phi = U(x) \phi_0 U^\dagger(x)\label{eq11}
\end{equation}
with arbitrary unitary matrix $U(x)$ is also a solution. Indeed
\[X_2(A_\mu(x),\phi) = X_2(U^\dagger(x) A_\mu(x) U(x), \phi_0) = 0\]
since Eq.~(\ref{eq10}) holds for any choice of $A_\mu$. In particular
one can choose $\phi_0$ diagonal. The generic $\phi_0$, diagonal and
constant, obeying Eq.~(\ref{eq10}) is (see appendix A)
\begin{equation}
\Phi^q_0 = \mathrm{diag }\, \left(
\overbrace{\frac{p}{N},\frac{p}{N},\ldots,\frac{p}{N}}^{q},
\overbrace{-\frac{q}{N},-\frac{q}{N},\ldots,-\frac{q}{N}}^p\right)
\label{eq12}
\end{equation}
where $p+q = N$, and $q = 1\ldots N-1$. If the solution is continuous,
it must be constant, since it cannot jump from one of the solutions in
Eq.~(\ref{eq12}) to another. So the generic solution to our initial
problem, which we call $\Phi$, is given by Eq.~(\ref{eq11}), where
$\phi_0$ can be any of the matrices $\Phi_0^q$ of Eq.~(\ref{eq12}).

Let us now show that, if $X_2=0$, then the terms $A_\mu \partial_\nu\Phi$
also vanish. Let us call $X_1(A_\mu,\Phi)$ such terms, we have
\begin{eqnarray}
X_1(A,\Phi)&=&{\rm Tr}\{\Phi (\partial_\mu A_\nu - \partial_\nu
A_\mu)\} +
{\rm Tr}\{\Phi\left[ [A_\mu,\Phi],\partial_\nu\Phi\right]\}
\nonumber \\
&& +{\rm Tr}\{\Phi\left[\partial_\mu\Phi, [A_\nu,\Phi]\right]\}
- {\rm Tr}\{\partial_\mu(\Phi A_\nu) - \partial_\nu(\Phi A_\mu)\}
  \nonumber \\
&=& {\rm Tr}\{\Phi\left[ [A_\mu,\Phi],\partial_\nu\Phi\right]\}
+{\rm Tr}\{\Phi\left[\partial_\mu\Phi, [A_\nu,\Phi]\right]\}
\nonumber \\
&& - {\rm Tr} \{\partial_\mu\Phi A_\nu - \partial_\nu\Phi A_\mu\}
\label{eq13}
\end{eqnarray}
Now~\cite{9}
\[ \partial_\mu\Phi = \partial_\mu (U\Phi_0 U^{\dagger}) =
[\Omega_\mu,\Phi]\qquad \Omega_\mu = (\partial_\mu U) U^{\dagger}\]
and hence from Eq.~(\ref{eq13})
\begin{eqnarray}
X_1(A,\Phi) &=&
{\rm Tr}\left\{\Phi\left[ [A_\mu,\Phi],[\Omega_\nu,\Phi]\right] +
\Phi\left[[\Omega_\mu,\Phi], [A_\nu,\Phi]\right] \right. \nonumber \\
& & \left. - [\Omega_\mu,\Phi] A_\nu - [\Omega_\nu,\Phi] A_\mu\right\}
\label{eq14}
\end{eqnarray}
which can also be written
\[ X_1(A,\Phi) = X_2(A+\Omega,\Phi) - X_2(\Omega,\Phi)  - X_2(A,\Phi)
\]
Since Eq.~(\ref{eq10}) is valid for arbitrary $A_\mu$, one has:
\[ X_1(A,\Phi) = 0\qquad {\rm if}\qquad X_2(A,\Phi) = 0\]
Then
\begin{equation}
F_{\mu\nu} =
\Tr\left\{
\partial_\mu(\Phi A_\nu) - \partial_\nu(\Phi A_\mu)\right\}
- \frac{i}{g}\Tr\left\{
\Phi[ \partial_\mu \Phi, \partial_\nu \Phi]\right\}
\label{eq15}
\end{equation}
In the abelian projected gauge, where $\Phi=\Phi_0^q$,
\begin{equation}
F_{\mu\nu} =
\Tr\left\{
\partial_\mu(\Phi A_\nu) - \partial_\nu(\Phi A_\mu)\right\}
\label{eq18}
\end{equation}
On the other hand, the 't Hooft tensor for a pure gauge field
$\omega_\mu = - i \left(\partial_\mu U\right) U^\dagger$ is zero
\begin{equation}
0 = F_{\mu\nu}(\omega) =
\Tr\left\{
\partial_\mu(\Phi \omega_\nu) - \partial_\nu(\Phi \omega_\mu)\right\}
- \frac{i}{g}\Tr\left\{
\Phi[ \partial_\mu \Phi, \partial_\nu \Phi]\right\}
\label{eq16}
\end{equation}
and finally, by use of Eq.~(\ref{eq16}), Eq.~(\ref{eq15}) can be rewritten
\begin{equation}
F_{\mu\nu} =
\Tr\left\{
\partial_\mu(\Phi A_\nu) - \partial_\nu(\Phi A_\mu)\right\} -
\Tr\left\{
\partial_\mu(\Phi \omega_\nu) - \partial_\nu(\Phi \omega_\mu)\right\}
\label{eq17}
\end{equation}
showing that $F_{\mu\nu}$ obeys Bianchi identities.

In conclusion an ``abelian'' field strength $F^q_{\mu\nu}$ can be
defined for each of the fields $\Phi^q$ of Eq.~(\ref{eq12}), or any
other field related to them by a gauge transformation. The
normalization of the solution is fixed. The invariance group of
$\Phi_0^q$ is
\begin{equation} SU(q)\times SU(N-q)\times U(1)
\label{eq19}
\end{equation}
and $\Phi^q(x)$ belongs to the coset of $\Phi_0^q$
\begin{equation}
\Phi^q(x) = U(x) \Phi_0^q U^\dagger(x)
\label{eq20}
\end{equation}
$\Phi_0^q$ defines a symmetric subspace~\cite{10}, in the sense that
the full algebra of $SU(N)$ is the sum of the subalgebra $L_0$ of the
little group of $\Phi_0^q$ plus the complement $L'$ to the full algebra,
and
\begin{equation}
[L_0,L_0] \in L_0\qquad [L',L_0] \in L'\qquad [L',L'] \in L_0
\label{eq21}
\end{equation}
The main property of symmetric spaces is that any element of the group
$U$ can be uniquely split in the form
\begin{equation}
U = e^{i L'} e^{i L_0} \label{eq22}
\end{equation}
a property which will be used below.

Let us now discuss the abelian projection in the light of the above
results. Following Ref.~\cite{7}, let us consider a generic operator
$X$, that transforms covariantly under gauge transformations. $X$ can
be diagonalized by a gauge transformation:
\begin{equation}
X_D(x) = U_X(x) X(x) U_X^\dagger(x) 
\label{eq:gauge_X}
\end{equation}
For each $\Phi^q_0$ in Eq.~(\ref{eq12}), a field transforming in the
adjoint representation of $SU(N)$ can be defined:
\begin{equation}
\Phi^q(x) = U_X^\dagger(x) \Phi^q_0 U_X(x)
\end{equation}
These fields $\Phi^q$ can now be used to define $N-1$ gauge-invariant
field strength tensors:
\begin{equation}
F^q_{\mu\nu} = {\rm Tr}\left\{\Phi^q G_{\mu\nu}\right\} 
-\frac{i}{g}{\rm Tr }\left\{\Phi^q
[\partial_\mu \Phi^q, \partial_\nu\Phi^q] \right\}
\end{equation}
In the gauge where $X$ is diagonal, $\Phi^q(x)=\Phi^q_0$ and,
according to the above results, these tensors reduce to the abelian
form:
\begin{equation}
F^q_{\mu\nu} =  \partial_\mu \Tr \left\{ \Phi^q_0 A_\nu \right\} -
\partial_\nu \Tr \left\{ \Phi^q_0 A_\mu \right\}
\end{equation}

The diagonal matrices $\Phi_0^q$ form a complete set of diagonal
matrices, and hence
\begin{equation}
X_D(x) = \sum_{q=1}^{N-1} c_q(x)\Phi_0^q
\label{eq24}
\end{equation}
We denote by $\alpha^a$ the diagonal matrices associated to the
simple roots via: 
\[ 
\alpha^a = \alpha^a_i H_i,
\] 
If the generators of the Cartan subalgebra $H_i$ are written in the
standard basis:
\[
(H_m)_{ij} = \left(\sum_{k=1}^m \delta_{ik}\delta_{jk} -
m \, \delta_{i,m+1}\delta_{j,m+1}\right)\frac{1}{\sqrt{2m(m+1)}}
\]
or
\[ H_m = \mathrm{diag }\,(
\overbrace{1,1,1\ldots 1}^m,-m,0\ldots 0)\frac{1}{\sqrt{2m(m+1)}}\]
the matrices associated to the simple roots $\alpha^a$ are:
\begin{eqnarray}
\alpha^1 = \alpha^1_i H_i &=& \frac{1}{2}\,\mathrm{diag }\,(1,-1,0,\ldots,0)
\label{eqa1}\\
\alpha^2 = \alpha^2_i H_i &=& \frac{1}{2}\,\mathrm{diag }\,
(0,1,-1,0,\ldots,0)\nonumber\\
\vdots\nonumber\\
\alpha^a = \alpha^a_i H_i &=& \frac{1}{2}\,
\mathrm{diag }\,(0,0,\ldots,0,\overbrace{1,-1}^{a,a+1},0,\ldots,0)\nonumber
\end{eqnarray}
It is then trivial to check that
\begin{equation}
\Tr\left\{\Phi_0^a \alpha^b\right\} = \frac{1}{2}\delta^{ab}
\label{eq25}
\end{equation}
and therefore
\[
\Tr\left\{X_D(x) \alpha^a\right\} = \frac{1}{2} c_a(x) = X_D^a -
X_D^{a+1}
\]
For each point $x$ such that $c_a(x) = 0$, two eigenvalues of $X$
become degenerate: 
\[
X_D^a(x) - X_D^{a+1} = 0,
\]
and the gauge transformation $U_X$, Eq.~(\ref{eq:gauge_X}), becomes
singular. Such a singularity behaves as a mag\-ne\-tic char\-ge with
respect to the $U(1)$ group of eq.(18) for $q=a$.

Let $A_\mu^D$ be the diagonal part of the gauge field in the abelian
projected gauge~\cite{7}
\begin{equation}
A_\mu^D = \mathrm{diag } \left(a^1_\mu, \ldots, a^N_\mu\right)
\label{eq:thooft_pot}
\end{equation}
The diagonal matrix $A^D_\mu$ can be expanded in
the form:
\begin{equation}
A_\mu^D = 2 \sum \tilde a_\mu^i \alpha^i\label{eq26}
\end{equation}
The $N-1$ abelian photons $\tilde a_\mu^i$ coincide with the abelian
fields defined via the abelian projection:
\begin{eqnarray}
\Tr\left\{\Phi^q_0 A_\mu^D \right\} &=& \tilde a_\mu^q  \\
F^q_{\mu\nu} &=& \partial_\nu \tilde a_\mu^q - 
\partial_\mu \tilde a_\nu^q \nonumber
\label{eq27}
\end{eqnarray}
The singularities of the gauge transformation $U_X$ at $c_a(x) = 0$
are magnetic charges with respect to the abelian field $\tilde a_\mu^a$,
a result that will be used to construct magnetically charged operators
in the next Section.

\section{Construction of the disorder parameter.}

In the abelian projected representation, in which the operator $\Phi$ is
diagonal, the generic link $U_\mu(x)$ can be cast in the form
\begin{equation}
U_\mu(x) = V^{(i)}_\mu(x) D_\mu^{(i)}(x)\qquad i = 1\ldots N-1
\label{eq28}
\end{equation}
where $D^i = \exp(i c^i_\mu \alpha^i)$, $\alpha^i$ being the diagonal
matrix corresponding to the simple root $\alpha^i$.

Indeed since the algebra of the little group of $\phi^a$, $L_0$,
defines a symmetric space, $U_\mu$ can be uniquely split as $U_\mu =
e^{i L'} e^{i L_0}$, as in Eq.~(\ref{eq22}). $\alpha^i$ is an element
of $L_0$ and commutes with all the others, since $\alpha^i$ is
invariant under $e^{i L_0}$
\begin{equation}
\Tr\left\{\Phi^a \alpha^b\right\} =
\frac{1}{2}\delta^{ab} = \Tr\left\{ e^{iL_0}\Phi^a e^
{-iL_0}\alpha^b\right\}
= \Tr\left\{ \Phi^a e^{-iL_0}\alpha^b e^{iL_0}\right\}
\label{eq29}
\end{equation}
$\alpha^i$ is the generator of the subgroup $U(1)$ of
Eq.~(\ref{eq19}). Therefore
\begin{equation}
U_\mu = e^{iL'} e^{i \tilde L_0} e^{i c^i_\mu \alpha^i}
\label{eq30}
\end{equation}
where $\tilde L_0$ is $L_0$ minus $\alpha^i$.
The plaquette $\Pi_{\mu\nu}(x)\equiv U_\mu(x) U_\nu(x+\hat\mu)
U_\mu^\dagger(x+\hat \nu) U^\dagger_\nu(x)$ can then be rewritten as
the product of matrices of the form in Eq.~(\ref{eq28})
\begin{eqnarray}
\Pi_{\mu\nu}(x) &=& V_\mu(x) D_\mu(x) V_\nu(x+\hat\mu) D_\nu(x+\hat\mu)
D_\mu^\dagger(x+\hat \nu)  V_\mu^\dagger(x+\hat \nu)
D^\dagger_\nu(x) V^\dagger_\nu(x) \nonumber \\
&=& V_\mu(x) \tilde V_\nu(x+\hat\mu) \tilde V_\mu^\dagger(x+\hat \nu)
\tilde V^\dagger_\nu(x) D_\mu(x) D_\nu(x+\hat\mu) D_\mu^\dagger(x+\hat \nu)
D^\dagger_\nu(x) \nonumber \\
&=&\tilde\Pi_{\mu\nu}(x) \Pi^0_{\mu\nu}(x)
\label{eq31}
\end{eqnarray}
An abelian $U(1)$ plaquette is thus defined. However an alternative
way of defining the abelian plaquette would be to operate the
separation of $\Pi_{\mu\nu}$ directly as done for the single links. It
is easy to see that the two definitions differ by terms ${\mathcal
O}(a^2)$. Indeed the second definition can be obtained by factorizing
$e^{iL_0}$ from the product $V^i \tilde V$ in Eq.~(\ref{eq31}). The
resulting $L_0$ would come from higher terms in the Baker Haussdorf
formula, which are ${\mathcal O}(a^2)$ and higher.

The lattice abelian projection is therefore intrinsically undefined by
terms of order ${\mathcal O}(a^2)$. A similar ambiguity comes out if the
abelian field is defined by $2\times2$ or $2\times1$ Wilson loops,
instead of the plaquette.

$U(1)$ monopoles are defined as in Ref.~\cite{DeGrand80}. The angle
$\theta_{\mu\nu}$ is defined by the equation
\begin{equation}
\Pi^0_{\mu\nu} = e^{i\theta_{\mu\nu}}
\label{eq32}
\end{equation}
The magnetic current defined as
\begin{eqnarray}
j_\mu &=& \Delta_\nu \theta^*_{\mu\nu} \nonumber \\
\theta^*_{\mu\nu} &=& \frac12 \epsilon_{\mu\nu\sigma\tau} \theta_{\rho\tau}
\end{eqnarray}
$j_\mu$
identically vanishes by the Bianchi identity. However, if the angle
$\theta_{\mu\nu}$ is defined modulo $2 \pi$:
\begin{eqnarray}
\theta_{\mu\nu} &=& \tilde \theta_{\mu\nu} + 2 \pi n_{\mu\nu},
-\pi \leq \theta_{\mu\nu} \leq \pi
\end{eqnarray}
then the magnetic current
\begin{eqnarray}
\tilde j_\mu &=& \Delta_\nu \tilde\theta^*_{\mu\nu}
\end{eqnarray}
can be different from zero and is conserved,
$\Delta_\mu \tilde j_\mu = 0$. The term proportional to
$n_{\mu\nu}$ counts the Dirac strings going through the plaquette,
which are invisible.

A monopole at a fixed time exists in an elementary spatial cube such
that one of the faces has $n_{ij}=1$, the others $n_{ij} = 0$. The
visible $\tilde \theta^*_{\mu\nu}$ has then a flux of $2\pi$, which is
balanced by the outgoing invisible string.

The operator $\mu$ which adds a monopole at the site $\vec y$ and time
$y^0$ to a generic configuration, can be constructed as follows
\begin{equation}
\mu = \exp\left\{-\beta\sum(\tilde S_{0i}(y_0) - S_{0i}(y_0))\right\}
\label{eq33}
\end{equation}
$S_{0i}$ are the terms of the action involving space-time $(0,i)$
loops, with a space link at $y_0$ and the others at $t\geq y_0$.  For
example for the Wilson action
\[ S_{0i} = \sum_{\vec n}\Pi_{0i}(y_0,\vec n)\]
We will recall the construction for Wilson action: the generalization to
actions
containing loops other than the plaquette, e.g. improved actions, is
straightforward.

In the favored abelian projection, according to Eq.~(\ref{eq30})
\begin{equation}
\Pi_{0i}(y_0,\vec n) = \tilde\Pi_{0i}(y_0,\vec n) \Pi^0_{0i}(y_0,\vec n)
\label{eq34}
\end{equation}
$\tilde S$ is obtained from $S$ by the following substitution in
$\Pi_{0i}(y_0,\vec n)$:
\begin{equation}
U_i(y_0,\vec n)\to U_i(y_0,\vec n) e^{i\alpha^a b_i(\vec n-\vec y)}
= U'_i(y_0,\vec n)
\label{eq35}
\end{equation}
where $b^\perp_i$ is the transverse component of the vector potential
generating at $\vec n$ a monopole sitting at $\vec y$, $\partial_i
b^\perp_i = 0$.
Classical gauge ambiguities in $b_i$ are contained in the longitudinal
part, and do not affect the definition Eq.~(\ref{eq35}).

The plaquette $\Pi_{0i}(y_0,\vec n)$ gets transformed to
$\Pi^\prime_{0i}(y_0,\vec n)$
\begin{equation}
\Pi^\prime_{0i}(y_0,\vec n)=
U_i(y_0,\vec n) e^{i\alpha^a b^\perp_i(\vec n - \vec y)}
U_0(y_0,\vec n+\hat i) U^\dagger_i(y_0+ 1,\vec n) U^\dagger_0(y_0,\vec n)
\label{eq36}
\end{equation}
this can be viewed as a change of $U^\dagger_i(y_0+ 1,\vec n)$:
\[ U^\dagger_i(y_0+ 1,\vec n) \to
U_0^\dagger(y_0,\vec n+\hat i) e^{i\alpha^a b^\perp_i(\vec n - \vec y)}
U_0(y_0,\vec n+\hat i)
U^\dagger_i(y_0+ 1,\vec n)\]
or
\begin{equation}
U_i(y_0+ 1,\vec n) \to U_i(y_0+ 1,\vec n)
U_0^\dagger(y_0,\vec n+\hat i)e^{-i\alpha^a b^\perp_i(\vec n - \vec y)}
U_0(y_0,\vec n+\hat i)\label{eq37}
\end{equation}
which is a multiplication of the link variable by an $SU(N)$ matrix
and it can be reabsorbed in a change of variables. However $U_i(y_0+
1,\vec n)$ also appears in the plaquettes $\Pi_{ij}(y_0+1, \vec n)$
and $\Pi_{0i}(y_0+1, \vec n)$. Up to terms ${\mathcal O}(a^2)$, the net
effect will be that in $\Pi_{ij}(y_0+1,\vec n)$,
\[
\theta^0_{ij} \to \theta^0_{ij} + \Delta_i b_j - \Delta_j b_i
\]
i.e. that a monopole has been added at $y_0+1$, and that a change like
the one in Eq.~(\ref{eq36}) is produced at time $y_0+1$.

By successive iteration of this procedure one eventually come at a
time $y^\prime_0$ where an anti-monopole of type $a$ is situated, and then
the procedure stops.

At $T=0$ the correlator $\langle\bar\mu(\vec x,t)\mu(\vec x,0)\rangle$
can be measured, and by cluster property
\[
\langle\bar\mu(\vec x,t)\mu(\vec x,0)\rangle
\mathop\simeq_{t\to\infty}
|\langle\mu\rangle|^2 + c e^{-M t}
\]
whence $|\langle\mu\rangle|$ can be extracted.

At $T\neq 0$ $\langle\mu\rangle$ is measured directly, and $C^*$ boundary
conditions in time are needed~\cite{1,3,4}.

By the appropriate choice of $b_i^\perp(\vec x-\vec y)$ in Eq.~(35)
a generic number of monopoles and anti-monopoles can be created at time
$y_0$.

In numerical simulations it is convenient to measure
\[
\rho = \frac{d}{d\beta}\log \langle\mu\rangle
\]
which is much less noisy, and in terms of which
\[ 
\langle\mu\rangle = \exp\left(\int_0^\beta \rho(x) dx\right)
\]
The statement that $\langle\mu\rangle\neq 0$ for $T< T_c$ in the
infinite volume limit corresponds to have $\rho$ volume independent
and finite at large volumes.  $\langle\mu\rangle = 0$, for $T> T_c$,
is obtained if
\begin{equation}
\rho \to - k L + k'\qquad (k>0)
\label{eq38}
\end{equation}
as the spatial size of the lattice, $L$, diverges. A behavior like
Eq.~(\ref{eq38}) is numerically easy to test, and means that
$\langle\mu\rangle$ is strictly zero in the thermodynamical limit. A
direct measurement of $\langle\mu\rangle$ would only produce a value
which is zero within (large) errors.

\section{Conclusions}
We have analyzed how to investigate dual superconductivity of the
vacuum in the confined phase of $SU(N)$ gauge theories for arbitrary
$N$, by deriving in detail the abelian projection, its symmetry
properties, and the construction of a disorder parameter. Numerical
simulations are in progress.
We plan to measure $\langle\mu\rangle$, or better $\rho$, as a function of
$\tilde \beta = \beta/N^2$ for different values of $N$. If the ideas
about $1/N$ apply $\rho N^2$ should be a universal function of
$\tilde \beta$, at sufficiently large $N$. Also the comparison of the
different choices of $\langle\mu\rangle$ and of different abelian
projections is of interest, on the way to understand the nature of dual
excitations of QCD.

This work is partially supported by MIUR, project ``Teoria e
Fenomenologia delle particelle elementari'', and by INTAS, Project
00-0110. B.L. is supported by the Marie Curie Fellowship
No. HPMF-CT-2001-01131. LDD thanks the Physics Department in Oxford
for hospitality during the final stage of this work.

\section{Appendix}
We want to show that the general diagonal, $x$ independent solution of
Eq.~(\ref{eq1}) has the form of Eq.~(\ref{eq10}). Let us call $H_i$
$(i=1\ldots N-1)$ the independent generators belonging to the Cartan
algebra of the group, $E_\alpha$ the generators belonging to the root
$\alpha$.

\noindent
The Lie algebra reads then
\begin{eqnarray}
[H_i,H_j] = 0 &&\qquad [H_i,E_\alpha ] = \alpha _i E_\alpha
\label{ar1} \\
\left[E_\alpha, E_\beta\right] = N_{\alpha,\beta} E_{\alpha+\beta} &&
\qquad [E_\alpha,E_{-\alpha}] = \alpha_i H_i \nonumber
\end{eqnarray}
In the fundamental representation
\begin{eqnarray}
\Tr\{H_i H_j\} = \frac{1}{2}\delta_{ij}
&\qquad& \Tr\{ E_\alpha^\dagger E_\alpha \} = \frac{1}{2}
\delta_{\alpha \beta} \nonumber \\
E^\dagger_\alpha = E_{-\alpha } 
&\qquad& \Tr\{H_i E_\alpha \} = 0 \nonumber
\end{eqnarray}
In general for a diagonal $\Phi$
\begin{equation}
\Phi = \sum c_i H_i
\label{eq40}
\end{equation}
\begin{equation}
A_\mu =   A^i_\mu H_i + A_\mu^\alpha E_\alpha
\label{eq41}
\end{equation}
We want to to solve Eq.~(\ref{eq10})
\begin{equation}
X_2(A_\mu,\phi) \equiv
{\rm Tr}\left(
\phi[A_\mu, A_\nu]\right) +
{\rm Tr}\left(\phi\left[[A_\mu,\phi],[A_\nu,\phi]\right]\right) = 0
\label{eq42}
\end{equation}
Using the multiplication law of the algebra, Eq.~(\ref{ar1}), and
Eq.~(\ref{eq41})
\begin{equation}
\left[A_\mu ,A_\nu \right] = A_\mu^\alpha  A^\beta_\nu \left[ E_\alpha,
E_\beta \right] + A_\mu^i  A^\beta_\nu \alpha_i E_\alpha
- A_\mu^\alpha  A^i_\nu \alpha_i E_\alpha
\end{equation}
The only contribution to $X_2$ comes from the first term of Eq.~(\ref{eq42})
and gives
\[
\frac{1}{2} c_i\alpha_i A_\mu^\alpha  A^{-\alpha}_\nu
\]
As for the second term of Eq.~(42), since
\[
\left[\hph, A_\mu \right] =
c_i A_\mu^\alpha [H_i, E_\alpha ] = c_i\alpha_i A_\mu^\alpha E_\alpha
\]
one has:
\[
\left[ [\hph, A_\mu ], [\hph, A_\nu ]
\right] = A_\mu^\alpha A_\nu^\beta
c_i\alpha_ic_j\beta_j[E_\alpha,E_\beta]
\]
and the second contribution to $X_2$ is
\[- \frac{1}{2} A_\mu^\alpha  A^{-\alpha}_\nu (c_i \alpha_i)^3\]
In summary $X_2=0$ is equivalent to
\begin{equation}
c_i\alpha_i =  (c_i\alpha_i)^3 \qquad \forall \alpha
\label{eq44}
\end{equation}
By gauge transformations we can order the eigenvalues of $\Phi$ in
decreasing order
\begin{equation}
\Phi = \mathrm{diag }(\varphi_1\ldots \varphi_n)\qquad \varphi_1 \geq
\varphi_2\geq\ldots
\varphi_n
\label{eq45}
\end{equation}
It is easy to find $N-1$ independent solutions of Eq.~(\ref{eq44}),
$\Phi^q = 2 \mu^q$ where $\mu^q$ are the fundamental weights
\begin{equation}
\mu^q = \frac{1}{2}
\left(
\overbrace{\frac{p}{N},\frac{p}{N},\ldots,\frac{p}{N}}^{q},
\overbrace{-\frac{q}{N},-\frac{q}{N},\ldots,-\frac{q}{N}}^p\right)
\label{eq46}
\end{equation}
with $q=1,2,\ldots,N-1$.

Indeed for the simple roots
\[ 2\Tr\left\{\mu^a\alpha^b\right\} = \delta^{ab}\]
A generic positive root $\alpha$ is the sum of simple roots, so that
\begin{eqnarray*}
2\Tr\left\{\mu^a \alpha\right\} = 1 && \hbox{if $\alpha$ contains
$\alpha^a$}\\
2\Tr\left\{\mu^a \alpha\right\} = 0 && \hbox{if $\alpha$ does not
contains
$\alpha^a$}
\end{eqnarray*}
The sign changes for negative roots. The $\mu$'s form  a complete set
for diagonal
operators. The general solution will be of the form
\[\Phi = 2\sum_A c^A\mu^A\]
\[ 2\Tr(\hph \alpha^A) = c^A = 0,\pm1\]
it is easy to see from Eq.~(\ref{eq46}) that
\begin{eqnarray*}
\Phi_1 - \Phi_2 &=& \frac{c_1}{N}\\
\Phi_2 - \Phi_3 &=& \frac{c_2}{N}\\
\ldots
\end{eqnarray*}
Since the eigenvalues are ordered all the coefficients $c^A$ must be non
negative.
if two $c$'s were different from zero, $c^i$, $c^j$, then for the root
$\alpha = \alpha^i + \alpha^{i+1}+\ldots \alpha^j$
we would have for the
scalar product
\[ 2 \Tr\left\{\Phi\alpha\right\} = 2\]
which is impossible. Therefore only one $c$ can be different from zero
and $\Phi^a$ provides the generic solution of Eq.~(\ref{eq10}).


\begin{thebibliography}{99}
\bibitem{1} A. Di Giacomo, B. Lucini, L. Montesi and G. Paffuti,
Phys. Rev. D61  034504 (2000).
\bibitem{2} A. Di Giacomo, B. Lucini, L. Montesi and G. Paffuti,
Phys. Rev.  D61  034505 (2000).
\bibitem{3} J.M. Carmona, M. D'Elia, A. Di Giacomo, B. Lucini, G. Paffuti,
Phys. Rev. D64, 114507, (2001).
\bibitem{4} J.M. Carmona, M. D'Elia, A. Di Giacomo, B. Lucini,
Nucl. Phys. B (Proc. Suppl.) 106, 07, (2002).
\bibitem{5} G. 't~Hooft, Nucl. Phys. B72, 461 (1974).
\bibitem{6} G. Veneziano, Phys. Rev. B117, 519 (1976).
\bibitem{7} G. 't~Hooft, Nucl. Phys.  B190, 455 (1981).
\bibitem{8} G. 't~Hooft, Nucl. Phys.  B79, 276 (1974).
\bibitem{9} A. Di Giacomo, M. Mathur, Phys. Lett. B400, 129 (1997).
\bibitem{10} S. Weinberg, ``The Quantum Theory of Fields'', Vol. 2.
\bibitem{DeGrand80} T. DeGrand, D. Toussaint, Phys. Rev. D22, 2478
(1980).
\bibitem{11} A. Di Giacomo, G. Paffuti, Phys. Rev.   D56, 6816 (1997).
\end{thebibliography}
\end{document}